\documentclass[a4paper,14pt]{article}
\usepackage{graphicx}
\usepackage{amsmath}
\title{Final State Interaction Effect in Pure Annihilation $B_s \to \rho \rho $ Decay }
\author{Mohammad Rahim Talebtash\footnote{mtalebtash@students.semnan.ac.ir},
 Hossein Mehraban\footnote{hmehraban@semnan.ac.ir}\\
Physics Department, Semnan University\\
P.O.Box 35195-363, Semnan, Iran}
\begin{document}
\maketitle
\begin{abstract}
We analyzed the process of $B_s\rightarrow \rho \rho$ decay in QCD
factorization (QCDF) and final state interaction (FSI) effects. In QCDF
for this decay we have only the annihilation graph and we expected
small Branching ratio. Then we considered FSI effect as a sizable
correction where the intermediate states are $\pi^0\pi^0$, $\pi^+\pi^-$,  $K^0\bar{K^0}$ and
$K^+K^-$ mesons. To consider the amplitudes of
these intermediate states, QCDF approach was used. The
experimental branching ratio of $B_s\rightarrow \rho\rho$ is less
than $3.20\times 10^{-4}$ and our result is $1.08 \times 10^{-9}$ and
$3.29 \times10^{-4}$ from QCDF and FSI, respectively.
\end{abstract}

\section{Introduction} 
Final state interaction (FSI) effects in B decay was expected to play only the role of a small correction to the standard description in short distance amplitude. In factorization approach, the amplitude of a B decay mode which is describe the short distance contributions, consists of 1) the usual factorization amplitude of color-allowed and color-suppression, 2) the annihilation topology (w-exchanged or w-annihilation)~\cite{1}. In pure annihilation B decay mode, the theoretical amplitude is often too small  in comparison to expected date. In this decay mode FSI effects may  play an important role. Where after a weak decay, the intermediate state particles re-scatter into the final particles through a nonperturbative strong interaction. The nonperturbative nature of FSI effects makes it  difficult to
study in systematic way so some different mechanisms of the rescattering effects have been considered~\cite{59,08}. To analyze a B-meson
decay through FSI,  it is important to understand the structure of the intermediate multiparticle states. One can treats FSI as the soft rescattering processes of intermediate two-body hadronic states e.g.  $B_s \to  K^0 \bar{K^0} \rightarrow \rho\rho$ and omit the other intermediate multi-body states, where after weak decay of B-meson to two light mesons, they rescatterd to two new mesons through nonperturbative  strong interaction. The hadronic loop level (HLL) is used in the strong interaction process  where it is obtained from the effective chiral Lagrangian~\cite{59,08}.\\
 Our result for QCDF approach was $1.08 \times 10^{-9}$ where the
leading order results for coefficients $C_i$ was used and the correction terms was omitted. Experimental result is less than $3.20 \times 10^{-4}$ ~\cite{EXP}. since, the results from QCDF approach is very small, the FSI effect can may give a sizeable correction where the intermediate states are $\pi^0\pi^0$, $\pi^+\pi^-$,  $K^0\bar{K^0}$ and
$K^+K^-$ mesons. We calculated the $B_s \to \rho \rho$ decay according to the HLL method. In this case, the branching ratio is $3.29 \times 10^{-4}$.\\
This paper is organized as follows. In section 2 we present the QCDF approach and calculated the amplitudes of the main decay and the intermediate states through QCDF approach. Then, in section 3 we present the FSI effects and calculated the amplitude of $B_s \to \rho \rho $ decay from three possible intermediate states. In
section 4 we give the numerical results, and in the last section we have a summery.

\section{Weak amplitude of the pure annihilation B decays }
To calculate the amplitudes of the pure annihilation B decay modes we use the QCD factorization method where we just consider the  annihilation topology. We consider b-quark decay and use the convention that $M_1 (M_2)$ meson contain an anti-quark (quark) from the weak vertex with longitudinal momentum fraction $\bar{y} (x)$  where  $M_1$ and $M_2$ are the final mesons ~\cite{brqcd}.
The weak annihilation contributions to the decay $B \to M_1M_2$ can be described in terms of the
building blocks $b_i$ and $b_{i,EW}$
\begin{eqnarray}
M(B \rightarrow M_1M_2) = -i 
\frac{G_F}{\sqrt{2}} \sum_{p=u,c}\lambda_p f_B f_{M_1}f_{M_1} \sum_{i}(d_ib_i+d_i'b_{i,EW}).
\end{eqnarray}
Where $\lambda_p = V_{pb} V_{pq}^*$ with $q = d, s$  and the building blocks have the expressions~\cite{bi}
\begin{eqnarray}
b_1& = &\frac{C_F}{N_C^2}C_1A_1^i,\nonumber\\
b_2& = &\frac{C_F}{N_C^2}C_2A_1^i,\nonumber\\
b_3& = &\frac{C_F}{N_C^2}[C_3A_1^i+C_5(A_3^i+A_3^f)+N_cC_6A_3^f],\nonumber\\
b_4& = &\frac{C_F}{N_C^2}[C_4A_1^i+C_6A_2^f],\nonumber\\
b_{3,EW}& = &\frac{C_F}{N_C^2}[C_9A_1^i+C_7(A_3^i+A_3^f)+N_cC_8A_3^f],\nonumber\\
b_{4,EW}& = &\frac{C_F}{N_C^2}[C_{10}A_1^i+C_8A_2^f],
\end{eqnarray}
The subscripts 1, 2 and 3 of $A_n^{i,f}$ denote the annihilation amplitudes induced from  $(V-A)(V-A)$, $(V-A)(V+A)$ and $(S-P)(S+P)$ operators, respectively, and the superscripts $i$ and $f$ refer to gluon emission
from the initial and final-state quarks, respectively and  shown in Fig.1 and given by~\cite{bi}
\begin{figure}
\centering
\includegraphics [scale=.6]{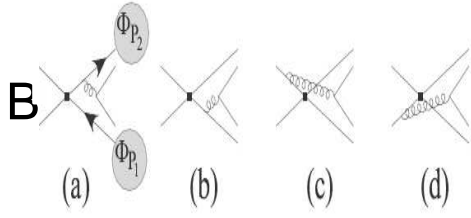}
 \caption{Annihilation correction to $B \to P_1 P_3$, where (a)and (b) correspond to $A_1^f$, while(c) and (d) give rise to $A_1^i$.}
\end{figure}
\begin{eqnarray}
A_1^i & =& \pi \alpha_s \int_0^1 dxdy\big{(}\Phi_{M_2}(x)\Phi_{M_1}(y)[\frac{1}{y(1-x\bar{y})}+\frac{1}{\bar{x}^2y}]+r_\chi^{M_1}r_\chi^{M_2}\Phi_{m_2}(x)\Phi_{m_1}(y)\frac{2}{\bar{x}y}\big{)},\nonumber\\
A_1^f & =&0 ,\nonumber\\
A_1^i & =& \pi \alpha_s \int_0^1 dxdy \big{(}\Phi_{M_2}(x)\Phi_{M_1}(y)[\frac{1}{\bar{x}(1-x\bar{y})}+\frac{1}{\bar{x}y^2}]+r_\chi^{M_1}r_\chi^{M_2}\Phi_{m_2}(x)\Phi_{m_1}(y)\frac{2}{\bar{x}y}\big{)},\nonumber\\
A_2^f & = &0 ,\nonumber\\
A_3^i& = &\pi \alpha_s \int_0^1 dxdy \big{(}r_\chi^{M_1}\Phi_{M_2}(x)\Phi_{m_1}(y)\frac{2\bar{y}}{\bar{x}y(1-x\bar{y})}-r_\chi^{M_2}\Phi_{M_1}(y)\Phi_{m_2}(x)\frac{2x}{\bar{x}y(1-x\bar{y})}\big{)},\nonumber\\
A_3^f& = &\pi \alpha_s \int_0^1 dxdy \big{(} r_\chi^{M_1}\Phi_{M_2}(x)\Phi_{m_1}(y)\frac{2(1+\bar{x})}{\bar{x}^2y}-r_\chi^{M_2}\Phi_{M_1}(y)\Phi_{m_2}(x)\frac{2(1+y)}{\bar{x}y^2}\big{)}. 
\end{eqnarray}

When all the basic blokes equation are solved, we found that weak annihilation kernels exhibit endpoint divergent~\cite{bi}:
\begin{eqnarray} \label{X}
X_A&=&\int_0^1 \frac{dy}{y}.
\end{eqnarray}
 Since the treatment of this logarithmically divergence is model depended, sub leading power corrections generally can be studied only in a phenomenological way. While the endpoint divergence is regulated in Perturbative QCD approach by introducing the parton's transverse momentum, it is parametrized  in QCD factorization by modifing $y \to y+\epsilon$ whith $\epsilon = O(\lambda_{QCD}/m_B)$ ~\cite{bi,cheng05}, so we replace Eq.~\eqref{X}‬ by:
\begin{eqnarray}
X_A&=&\int_0^1 \frac{dy}{y+\epsilon}=\ln \frac{m_B}{\lambda_h}(1+\rho_Ae^{i\Phi_A}).
\end{eqnarray}
Different $X_A$ are allowed for four cases: PP, PV, VP and VV where P(V) is a final meson by pseudoscalar (vector) polarization. for VV case, in ~\cite{be03}, by evaluating the convolution  integrals with asymptotic distribution amplitudes $\Phi(x)=\Phi_\parallel(x)=6x\bar{x}$, $\Phi_p=1$, and $\Phi_v(x)=3(x-\bar{x})$, we find the simple expressions
\begin{eqnarray}
A_1^i &\simeq& A_2^i =2 \pi \alpha_s (9(X_A-4+\frac{\pi^2}{3})+(r_\perp^V)^2(X_A-2)^2),\nonumber\\
A_3^i&=&0,\nonumber\\
A_3^f&=&-36\pi \alpha_s r_\chi (2X_A^2-5x_A+2).
\end{eqnarray}
Now we can calculate the weak amplitude of the $B_s \to \rho \rho$ decay and the intermediate state.
According to the annihilation diagrams of the  $B_s\to \rho \rho$ decay which is given in Fig.2  the pure annihilation amplitude is given by
\begin{eqnarray} M(B \rightarrow \rho \rho) = -i \frac{G_F}{\sqrt{2}} f_{B_s} f_{\rho}^2 V_{tb}V^*_{td}(2b_4-b_{4,EW}).
\end{eqnarray} 
\begin{figure}[h] \centering
\includegraphics[scale=.4]{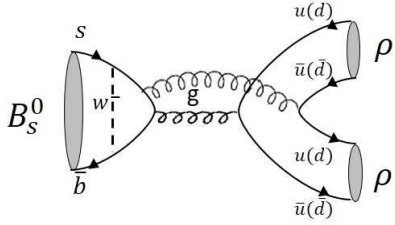}
 \caption{Feynman diagram for $B_s \rightarrow \rho \rho$ decay }
\end{figure}

\section{weak amplitude of the intermediate states }
To consider the FSI effects in $B_s \to \rho \rho$ decay, we must extract the accessible intermediate states and  calculate the  weak amplitude of them. According to the Fig.3, by considering  the $u \bar{u}$ part of the $\rho$ mesons while  tow intermediate mesons  and final state mesons,  exchange the same quark ( u-quark ),  $\pi^0$ and $\pi^0$ meson  can be produced for the intermediate state via exchange $\pi^0 (\omega)$ meson. Likewise when two intermediate mesons exchange d-quark (s-quark) and two final state mesons exchange u-quark, $\pi^+\pi^-$ ($K^{(*)+}$  $K^{(*)-}$) mesons can be produced for the intermediate state via exchange $\pi^+$ ( $K^{(*)0}$) mesons. And by considering the  $d \bar{d}$ part of the $\rho$ mesons while the  two intermediate mesons exchange d-quark (u-quark or s-quark ) and two final state mesons exchange d-quark,  $\pi^0\pi^0$ (  $\pi^+\pi^-$ or $K^{(*)0} \bar{K}^{(*)0}$ ) mesons  can be produced for the intermediate state via exchange $\pi^0 (\omega)$  ( $\pi^+$ or $K^{(*)0}$ ) meson.
\begin{figure}[h]
\centering
\includegraphics[scale=.5]{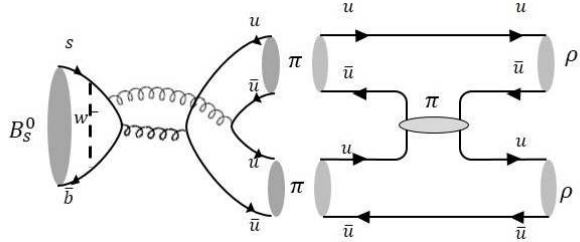}
 \caption{Quark level diagram for $B_s \to \pi \pi \to \rho \rho $.}
\end{figure}
Now that the intermediate states obtained, we can calculate the weak amplitude of these intermediate states which produced in  $B_s \to m_1 m_2$ decay modes where $ m_1$ and $ m_2$ are the intermediate state mesons.
 In tow case,  $\pi^0$  $\pi^0$ and  $\pi^+\pi^-$, we calculate the amplitude is similar to the $B \to \rho \rho$ decay mode since these decay modes are pure annihilation. So according to sec.2, we have
 \begin{eqnarray}
M(Bs \to \pi^0\pi^0(\pi^+\pi^-)) =-i G_f f_{B_s} f_{\pi}^2(V_{ub}V_{us}^*b_1+2V_{tb}V_{ts}^*b_4).
\end{eqnarray}
 But in the other cases, the  color-allowed and color-suppression topology are allowed and  we must consider  the usual
factorization approach ~\cite{be03,c} to calculate the amplitude. So we have
\begin{eqnarray}
M(B_s \rightarrow K^{0} \bar{K}^{0}) &=& -i
G_F V_{tb} V^*_{td}\bigg\lbrace f_{K} F_0^{B_sK^0}(M^2_{B_s}-M^2_K)  a_4\nonumber \\
&&+f_{B_s} f_K^2 (2b_4)\bigg\rbrace,
\end{eqnarray}
 where the coefficients $a_4$ correspond to the
penguin topology and is defined as:
\begin{eqnarray}
a_4 &=& C_4 + \frac{1}{N_c} C_{3},
\end{eqnarray}
In the QCD factorization amplitude, all terms are not expected to be equally large. The color-allowed and color-suppression topology ($a_i$ terms) which invole form factors are dominate  and the annihilation topology ($b_i$ terms) can be neglected~\cite{10}.\\
Likewise we have:
\begin{eqnarray}
M(B_s \rightarrow K^{0*} \bar{K}^{0*}) &=& -i
\frac{G_F}{\sqrt{2}}\bigg\lbrace f_{K^*} m_{k^*}
 [(\epsilon_1^*.\epsilon_2^*)(m_{B_s}^2+m_{K^*}^2)A_1^{BK^*}(m_{K^*}^2)\nonumber \\
&& -(\epsilon_1.p_B)(\epsilon_2.p_B)\frac{2A_2^{Bk^*}(m_{k^*})}{(m_B^2+m_{K^*}^2)}]
 a_4 V_{tb} V^*_{td} \nonumber \\
&&-f_{B_s} f_{K^*}^2 V_{tb} V^*_{td} (2b_2)\bigg\rbrace,
\end{eqnarray}
\begin{eqnarray}
M(B_s \rightarrow K^+ K^-) &=& -i
G_F V_{ub} V_{us}^*\lbrace f_K F_0^{B_sK}(M^2_{B_s}-M^2_K)  a_1\rbrace \nonumber \\
&&+i G_F f_{B_s} f_K^2\lbrace V_{ub} V^*_{us} b_1+  V_{tb} V^*_{ts} (b_4+b_{4,Ew})\rbrace,
\end{eqnarray}
\begin{eqnarray}
M(B_s \rightarrow K^{0*} \bar{K}^{0*}) &=& -i
G_F\lbrace f_{K^*} m_{k^*}
 [(\epsilon_1^*.\epsilon_2^*)(m_{B_s}^2+m_{K^*}^2)A_1^{BK^*}(m_{K^*}^2)\nonumber \\
&& -(\epsilon_1.p_B)(\epsilon_2.p_B)\frac{2A_2^{Bk^*}(m_{k^*})}{(m_B^2+m_{K^*}^2)}]
  V_{ub} V^*_{ud} a_1 \rbrace \nonumber \\
&&+i G_f f_{B_s} f_{K^*}^2 \lbrace  V_{ub} V^*_{ud} b_1+ V_{tb} V^*_{td} (b_4-b_{4,EW})\rbrace.
\end{eqnarray}

\section{The one particle exchange method for FSI}
At the quark level, final state re-scattering can occur through quark exchange and quark annihilation. The quark  level diagram for $B \to \rho \rho $ decay is shown in Fig.3. This decay has only quark annihilation mode, since the final mesons ( $ \rho$ ) have the same flavour quark-antiquark.
In practice, it is extremely difficult to calculate the FSI effects, but at the hadronic level formulated as re-scattering processes with s-channel resonances and one particle exchange in the t-channel. S-channel resonant FSI effects in  $B \to \rho \rho $ decay is expected to be vanished because of the lack of the existence of resonances. Therefore, one can model FSI effects as re-scattering processes of two body intermediate state with one particle exchange in the t-channel and compute the absorptive part via the optical theorem~\cite{59}. 
\begin{figure}[h]
\centering {\includegraphics[scale=.6]{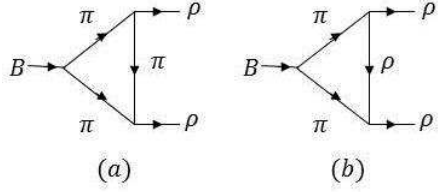}}
 \caption{t-channel contributions to final state interaction in  $B\to \pi \pi \to \rho \rho $ decay due to one particle exchange.}
\end{figure}
So, according to the hadronic loop level (HLL) diagram, shown in Fig.4, the absorbtive part of the amplitude calculated with the following formula
\begin{eqnarray} 
&&Abs \big ( B_s(p_B)\rightarrow \pi(p_1)\pi(p_2) \rightarrow \rho (p_3)\rho(p_4) \big) =\nonumber\\
 & &\hspace{3cm}  \int_{-1}^1 \frac{|\overrightarrow{p_1}|d(\cos
\theta )}{16 \pi m_B} A\big(B_s \rightarrow \pi (p_1)\pi(p_2)\big)\nonumber \\
& &\hspace{3.4cm}\times G\big(\pi (p_1)\pi(p_2)) \rightarrow \rho(p_3)\rho(p_4)\big),
\end{eqnarray}
where the $A(B_s \rightarrow \pi \pi)$ is the amplitude of the decay of the  B meson to the intermediate state and $G( \pi \pi \rightarrow \rho \rho)$ involves the hadronic vertices  factor, which are defined  as
\begin{eqnarray}
<\pi(p_3)\rho(p_2,\epsilon_2)|i\pounds|\pi(p_1)>&=&-ig_{\pi\pi\phi} \varepsilon_2.(p_1+p_3),\nonumber \\ 
<\omega(p_3,\varepsilon_3)\rho(p_2,\epsilon_2)|i\pounds|\pi(p_1)>&=&-ig_{\omega \pi\rho}
\varepsilon_
{\mu\nu\alpha\beta} \varepsilon_2^\mu \varepsilon_3^{*\nu} p_1^\alpha p_2^\beta \nonumber. \\
\end{eqnarray}
The magnitude of the effective couplings can be extracted from experimental~\cite{g}.  As an example, the effective coupling $g_{\pi\pi\rho}$ is relevant to the  $\rho \to \pi\pi$ process shown in Fig.5.\\
\begin{figure}[h]
\centering {\includegraphics[scale=.4]{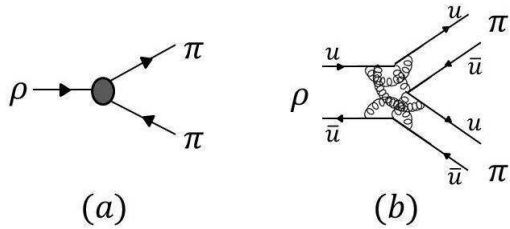}}
\caption{The effective coupling vertex on (a) hadronic, (b) quark level.}
\end{figure}
So, for diagram (a) in Fig.4 the absorptive part of the amplitude of the $B\to \pi\pi \to \rho \rho$ proses where $\pi $ meson is exchanged particle at t-channel is given by
\begin{eqnarray}
Abs(4a) &=& \int_{-1}^1\frac{|\overrightarrow{p_1}| d( \cos
\theta)}{16\pi m_{B_s}} M(B_s \to \pi\pi)
g_{\pi\pi\rho} (2\epsilon_2 . p_1) \nonumber \\
 & &\times g_{\pi\pi\rho} (2\epsilon_4 .p_2)\times\frac{F^2(q^2,m_\pi^2)}{q^2-m_\pi^2} \nonumber \\
&=&  \frac{g_{\pi\pi\rho}^2}{4 \pi m_{B_s}}  M(B_s \to \pi\pi )
\int_{-1}^1|\overrightarrow{p_1}|d(\cos \theta )
\frac{F^2(q^2,m_{\pi}^2)}{q^2-m_{\pi}^2}H_1,\end{eqnarray}
where the $\theta$ is the angle between $\textbf{p}_1$ and $\textbf{p}_3$ and $\textbf{q}=\textbf{p}_1-\textbf{p}_3$ is the four momentum of the exchanged K meson and
\begin{eqnarray}
H_1&=&(\varepsilon_3.p_1)(\varepsilon_4.p_2), \nonumber \\
q^2&=&m_1^2+m_3^2-2E_1E_3+2|\overrightarrow{p_1}||\overrightarrow{p_3}|\cos
\theta.
\end{eqnarray}
$F(q^2,m_q^2)$ is a form factor or a cut-off which is introduced to that hadronic vertices takes care of the off-shell effect of the exchanged particle.  Here we have followed  ~\cite{59} while in ~\cite{fajfer} different form  is used.  So we have 
\begin{eqnarray}
F(q^2,m_q^2)&=&\Big{(}\frac{\Lambda^2 -m_q^2}{\Lambda^2 -q^2} \Big{)}^2.
\end{eqnarray}
The parameter $\Lambda$ is the off-shellnes compensating in  function $F(q^2,m_q^2)$ which is not an universal parameter, but is should be near the mass of the mesons involved in the effective coupling. \\
Likewise, for the diagram (b) in Fig.4, the absorptive part for the $B\to \pi \pi \to \rho \rho$ proses where $\omega$ meson is exchanged particle at t-channel is given by
\begin{eqnarray}
Abs(4b)&=&\int_{-1}^1\frac{|\overrightarrow{p_1}|d(\cos
\theta)}{16\pi m_{B_s}}M(B_s\to \pi\pi)
(-i)g_{\omega\pi \rho}\varepsilon_{\mu\nu\alpha\beta}\varepsilon_3^\mu\varepsilon_q^{*\nu}
 p_1^\alpha p_3^\beta \nonumber \\
& &\times (-i)g_{\omega\pi \rho}\varepsilon_{\rho\sigma\lambda\eta}\varepsilon_4^\rho\varepsilon_q^{*\sigma}
 p_2^\lambda p_4^\eta  \, \, \frac{F^2(q^2,m_{\omega}^2)}{q^2-m_\omega^2} \nonumber \\
&=& \frac{g_{\omega\pi \rho}^2}{16\pi m_{B_s}}M(B_s\to \pi\pi)
\int_{-1}^1|\overrightarrow{p_1}|d(\cos \theta)
\frac{F^2(q^2,m_\omega^2)}{q^2-m_\omega^2}H_2,
\end{eqnarray}
where
\begin{eqnarray}
H_2&=&-2\big[(p_1.p_4)(p_2.p_3)-(p_1.p_2)(p_3.p_4)\big]+p_3^0\big[p_4^0(p_1.p_2)\nonumber \\
&&-(p_2^0-|\overrightarrow{p_2}|)(p_1.p_4)\big]+
(p_1^0-|\overrightarrow{p_1}|) \nonumber \\
&&\times\big[(p_2^0-|\overrightarrow{p_2}|)(p_3.p_4)-p_4^0(p_2.p_3)\big].
\end{eqnarray}
As the bridge between the dispersive part of the FSI amplitude and the absorptive part, the dispersion relation is
\begin{eqnarray}
Dis(m_B^2)&=&\frac{1}{\pi}\int_s^\infty \frac{Abs_a(s')+Abs_b(s')}{s'-m_B^2}ds',
\end{eqnarray} 
where $s'$ is the square of the momentum carried by the exchanged particle and $s$
is the threshold of intermediate states, in this case $s \sim m_B^2$
.\\
Finally the amplitude of the $B_s\to \pi\pi \to \rho\rho$ decay via HLL diagram is 
\begin{eqnarray}
A(B_s\to \pi\pi \to \rho\rho)&=&Abs(4a)+Abs(4b)+Dis(4).
\end{eqnarray}
Likewise, we can calculate the FSI effects comes from other intermediate states ($K^{(*)+}K^{(*)-}$ and ($K^{(*)0}K^{(*)0}$) by replacing  $\pi \pi $  with these new intermediate states in above formula.

\section{Numerical Results}
In this paper we used Wilson coefficients $c_i$ in leading order
(LO) at $\mu=m_b$ which are given by ~\cite{c}
\begin{tabbing}
\hspace{4cm}\=\kill
$ c_1=1.114$,\> $c_2=-0.308$,\\
 $c_3=0.014$,\>$c_4=-0.030$, \\
$c_5=0.009$, \>$c_6=-0.038$, \\
$c_7=-3.4\times10^{-4}$, \>$c_8=3.7\times10^{-4}$, \\
$c_9=-0.01$, \> $c_{10}=0.002$.
\end{tabbing}
The elements of the CKM matrix can be parametrized by three mixing angles $A, \lambda,\rho$ ~\cite{ckm} and a CP-violating phase $\eta$
\begin{tabbing}
\hspace{4cm}\=\hspace{4 cm}\=\kill
$V_{ud}=1-\lambda^2/2$,\>$ V_{us}=\lambda$, \> $V_{ub}=A\lambda^3(\rho-i\eta)$,\\
 $V_{cd}=-\lambda$,\> $ V_{cs}=1-\lambda^2/2$, \>$V_{cb}=A\lambda^2$, \\
 $V_{td}=A\lambda^3(1-\rho-i\eta)$,\> $V_{ts}=-A\lambda^2$,\> $ V_{tb}=1$.
\end{tabbing}
The results for the Wolfenstein parameters are
\begin{tabbing}
\hspace{4cm}\=\kill
$ \lambda=0.2257\pm0.001$, \> $A=0.814\pm0.02$,\\
 $ \bar{\rho}=0.135\pm0.023$, \>$\bar{\eta}=0.349\pm0.016$,
\end{tabbing}
and we use the central values of the Wolfenstein parameters and obtain
\begin{tabbing}
\hspace{4cm}\=\hspace{4 cm}\=\kill
$V_{ud}=0.9745$,\>$ V_{us}=0.2257$, \> $ V_{ub}=0.0013-0.0033i$,\\
 $V_{cd}=-0.2257$,\> $ V_{cs}=0.9745$, \>$V_{cb}=0.0415$, \\
 $V_{td}=0.0081-0.0033i$,\> $V_{ts}=0.0415$,\> $ V_{tb}=1$.
\end{tabbing}
For endpoint parametrizing in  QCD Factorization approach according to the polarization of the final mesons we give:
$\rho=1$,  $ \Lambda=0.5$, $\Phi_A=-40^0 (VV)$ , $\Phi_A=20^0 (PV)$, $\Phi_A=-55^0 (PP)$ ~\cite{cheng05}.\\
\\
The  mass of the mesons and decay constants are given in unit of GeV:\\
 $ m_B=5.28$, $m_K=0.49$ , $ m_{K^*}=0.89 $, $m_\rho=0.775$, $m_{\pi}=0.139$,
 $m_{\omega}=0.783$, $f_{B_s}=0.230$,   $f_k=0.16$, $f_{k^*}=0.214$,
 $f_{\pi}=0.133$, $f_{\rho}=0.216$,
  $f_{K^*}^\perp=0.175$,
$F_0^{B_sK}=0.26$, $A_1^{B_sK^*}=0.29$, $A_2^{B_sK^*}=0.26$,
.~\cite{1,g3}\\
\\
The other input parameters used are given by:\\
$g_{KK\rho}=g_{K^*K^*\rho}=3.025$,
$g_{\omega\pi\rho}=5.89$,$r_\chi^K=1.09$,$r_\chi^{K^*}=0.29$,~\cite{g3,g1,g2}\\
\begin{table}[h]
\caption{The branching ratio of $B \to \phi \phi $ decay with
$\eta =0.5 \sim  1 $. (in units of $10^{-4}$).} \centering
\begin{tabular}{ccccccccc}
  \hline
  $\eta$ & 0.5 & 0.6 & 0.7 & 0.8 & 0.9 & 1 & EXP.~\cite{EXP} \\
  \hline
  BR & 0.23 & 0.47 & 0.88 & 1.42 & 2.22 & 3.29 & $<$3.20 \\
  \hline
\end{tabular}
\end{table}
\\
 We calculated  the branching ratio for the QCD Factorization method as
$ 1.08 \times 10^{-9}$ which is very small compared to the
experimental result .Within FSI
 the branching ratio is shown in
table 2 and if $\eta=1 $ selected, the branching ratio is
 $ 3.22 \times 10^{-4}$ which is near to experimental result.

\section{summery}
We analyzed  the $B \to \rho
\rho $ decay in  QCD Factorization approach and then we added the final state interaction effects. In QCD Factorization approach we have just weak  annihilation topology and as we expected we have obtained a small branching ration $1.08 \times 10^{-9}$ while after considering the FSI effect we have obtained  $3.29 \times10^{-4}$ which is near to upper bound of the  experimental value which is  $3.20 \times 10^{-4}$~\cite{EXP}. The main
phenomenological parameter in FSI effects is $ \eta $ which is determined  from the measured ratios. Its value in form factor is expected to be of the order of unity. In this work we have considered  $\eta=0.5 \sim 1$  and the best result obtained by $\eta=1$.

\end{document}